\documentclass[%
 reprint,
 amsmath,amssymb,
 aps,
]{revtex4-2}
\usepackage{graphicx}% Include figure files
\usepackage{dcolumn}% Align table columns on decimal point
\usepackage{bm}% bold math
\begin{document}

\preprint{APS/123-QED}

\title{Dewetting of Thin Lubricating Films Underneath Aqueous Drops on Slippery Surfaces}

\author{Bidisha Bhatt, Shivam Gupta, Meenaxi Sharma}

\author{Krishnacharya Khare}

\email{kcharya@iitk.ac.in}

\affiliation{Department of Physics, Indian Institute of Technology Kanpur, Kanpur-208016, INDIA}

\begin{abstract}
Stability of thin lubricating fluid coated slippery surfaces depends on the surface energy of the underlying solid surface. High energy solid surfaces, coated with thin lubricating oil, lead to the dewetting of the oil films upon depositing aqueous drops on it. The total surface energy, which is due to the long range and short range interactions, also predict the instability of thin lubricating films under the given condition. In this article, we present experimental study of dewetting of thin lubricating oil films sandwiched between hydrophilic solid surface and aqueous drops. Fluorescence imaging of lubricant film and wetting behavior of aqueous drops are used for the analysis. We find that the dewetting dynamics and the final pattern depend strongly on the thickness of the lubricating oil film. 
\end{abstract}

\maketitle

%\tableofcontents
%\section{\label{Introduction}Introduction\protect\\}

Liquid-repellent surfaces got attention over a century because the drops roll-off easily on these surfaces only tilting by few degrees and have broad technological applications in the field of biomedical devices and fuel transport \cite{barthlott1997purity, falde2016superhydrophobic, david2005non, jokinen2018superhydrophobic, lafuma2003superhydrophobic}. One approach to achieve liquid-repellent surfaces is ``Slippery liquid infused porous surfaces (SLIPS or slippery surfaces)" inspired from the ``Nepenthes" pitcher plant. Slippery surfaces have micro/nano-grooves, and these grooves are infiltrated with the low surface energy liquid. These are nearly physically and chemically smooth, shows no pinning at the three-phase contact line for all the low-and high-surface energy liquids \cite{lafuma2011slippery, wong2011bioinspired, smith2013droplet, schellenberger2015direct, daniel2017oleoplaning}. The wetting property of slippery surfaces is defined by depositing a liquid drop. There are two possibilities arises after drop deposition, either drop will float on a lubricating film or it sinks. The floating and the sinking behavior depends on the interfacial property between the interface of infiltrated liquid, test liquid, and substrate. Quere's et. al. derived condition for sinking and floating, which is the change in the interfacial energy per unit area of the system before and after the drop deposition \cite{lafuma2011slippery}. In the floating situation, the drop is in direct contact with the film and feels no pinning force at the contact line and the drop moves (float) easily on it \cite{wong2011bioinspired, schellenberger2015direct}, but when the drop sinks it comes in the direct contact with the substrate and feels pinning force (presence of the dust particles or impurities) at the contact line and results contact angle hysteresis. Only a few works are available in the direction of the sinking of the droplet on slippery surfaces. The previously reported works mainly show macroscopic behavior i.e. the dynamics (sinking dynamics) of the droplet over the slippery surfaces \cite{pant2016slipperiness, sharma2019sink, carlson2013short}, but the phenomena (rupturing of the lubricating film) happen underneath the droplet is unclear. 

Thin polymer film used as the lubricating fluid in slippery surfaces often becomes unstable (dewetting) on solid surfaces due to its interaction with the substrate and test liquid by long-range interaction, short-range polar interaction, and other molecular forces \cite{sharma1993relationship, sharma1993equilibrium}. Despite the large applications of the system when drop floats, instability pattern arises by rupturing of the film when the drop sinks, used in many applications like protein patterning, cell culture, contact lens production, and water harvesting \cite{yoon2008nanopatterning, colusso2019fabrication, telford2017functional, verma2011submicrometer, gentili2012applications, sachan2015hierarchical, thickett2011biomimetic, huang2005spontaneous, doi:10.1002/adma.201002768}.
 
\begin{figure*}[htbp]
	\centering
		\includegraphics[width=1.00\textwidth]{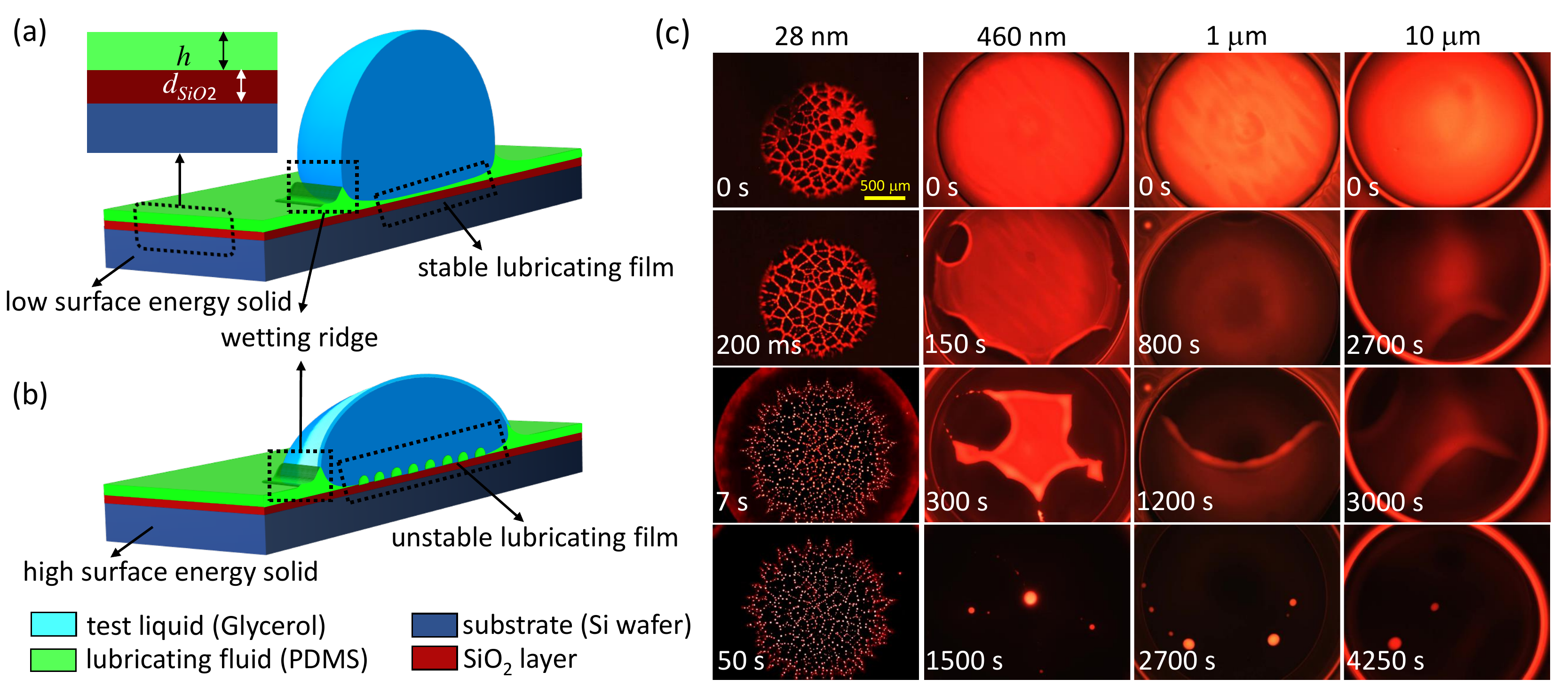}
	\caption{Cross section view of slippery surfaces (a) low energy substrate (stable underneath lubricating fluid), (b) high energy substrate (unstable underneath lubricating fluid ); (c) dewetting dynamics of the lubricating film thickness 28 nm, 460 nm, 1 $\mu$m and 10$\mu$m for high energy solid surface. Scale bar for all optical images is same.}
	\label{Fig:1}
\end{figure*}

In this article, we discuss the effect of the film thickness on morphological patterns obtained after the rupturing of thin lubricating films, due to the interfacial and intermolecular forces present in the system. Ultra-thin films with thickness $\leq$ 20 nm undergo spontaneous (spinodal) dewetting, wheares thicker films dewet via heterogeneous nucleation \cite{meredith2000combinatorial, jacobs1998thin}. During the growth of nucleated holes, the velocity of dewetting front depends on the the molecular interaction between the solid surface and the liquid which can be defined in terms of the ``slip length". Radii of growing holes follow the relation, D(t)$\sim$ {t}$^{\alpha}$, where $\alpha=1$ represents no-slip systems and $\alpha=2/3$ indicates the presence of slippage in the system. At the final stage of dewetting, holes start coalescing with neighboring ones forming liquid-ridges at the boundary of the holes. The breakup of a liquid-ridge follows Rayleigh-Plateau instability due to the presence of capillary waves at the liquid-ridge, which subsequently breaks into satellite and sub-satellite droplets \cite{tjahjadi1992satellite}. It has been found that the dewetting dynamics of thin liquid films depend on the wettability of substrate, film thickness, viscosity and surface tension of the liquid, and temperature \cite{baumchen2009slip, xie1998spinodal}. Most of the dewetting studies are performed with thin polymer films formed out of equilibrium in air where the film dewets either upon heating or bringing a suitable solvent \cite{beena2016solvent, xu2012dewetting, al2016good, verma2010enhanced}. Reiter et al. studied the dewetting of thin polymer films under aqueous environment due to solely long range force \cite {reiter1999thin, reiter1999strength}. In most of the dewetting studies, thickness of liquid films are kept about or below 100 nm \cite {sharma1998pattern, peschka2019signatures}, as even several hundred nanometer thick film dewet in few seconds time \cite {reiter2000enhanced}. Thin lubricating films of slippery surfaces provide an excellent alternative to study the dewetting of thick films up to 10 $\mu$m, which is reported here. Thin lubricating films, which are sandwitched between solid substrates and aqueous drops, may undergo dewetting if the solid substrate is hydrophilic (high surface energy). Lubricant films of different thicknesses (from few nanometers to few microns) dewet on such slippery surfaces which is characterized in terms of hole growth and apparent contact angle of aqueous drops. 

When aqueous drops are deposited on thin lubricating fluid coated solid surfaces, the lubricating film underneath the drops will be stable and unstable on low surface energy (hydrophobic) and high surface energy (hydrophilic) solid surfaces, respectively [Figs. 1(a) and 1(b)]. Unstable lubricating films undergo dewetting and ruptures into multiple droplets. Dewetting dynamics of underneath lubricating films is shown in Fig. 1(c) for different thickness of the film. Cleaned silicon wafer ($<$100$>$, University Wafer Inc.) having a native oxide layer of 1.7 nm and surface energy of 53.9 mJ/m$^2$ were used as the solid substrates. To prepare slippery surfaces, polydimethylsiloxane (PDMS) (sylgard 184, 10:1 ratio, viscosity $3500\,\mathrm{cSt}$, surface tension 21.2 mJ/m$^2$) was coated on silicon wafers at fixed $2000\,\mathrm{RPM}$, $10\,\mathrm{s}$ acceleration, and $100\,\mathrm{s}$ spin time. To vary the film thickness from 28 nm to $10\,\mu\mathrm{m}$, n-Heptane (Sigma Alderich) was used as a solvent for PDMS. Thickness of the lubricating films were measured using a surface profilometer (Veeco Dektak 6M Stylus) after crosslinking the films. For fluorescence imaging, oil miscible dye, Nile Red (Sigma Alderich, ), was mixed in the lubricating fluid. Glycerol drops ($\gamma_{G}=\,\mathrm{64\,mJ/m^2}, \rho_{G}=1.26\,\mathrm{g/mL}$, sigma) of 2 $\mu$l volume were deposited on the lubricating films and fluorescence microscopy and optical contact angle measurements were performed to investigate the dewetting of the underneath lubricating films. All the length scales in our system (drop size and film thickness) are kept below the capillary lengths to ignore any inertial effects. 

As shown in Fig. 1(c), for 28 nm thick lubricating film, multiple holes are nucleated in few ms time after the drop deposition. The hole nucleation starts at a place where a glycerol drop first touches the lubricating film (intial contact line of the glycerol drop). The nucleated holes grow in size with time and subsequently coalesce with each other. However, as the nucleated holes grow, the top glycerol drop spreads as it gets partially exposed to the hydrophilic solid surface. Upon complete dewetting, the lubricating film breaks into the micrometer sized small droplets ordered in polygon structure. Although, dewetting of the film completes in about 50 s time, the glycerol drop continues to spread on the dewetted hydrophilic solid surface. Dewetted PDMS drops are not visible in the region near to the contact line of the glycerol drop during spreading, which indicates that the drop it completely removes the lubricating fluid underneath the drop during spreading (see supplementary). Since the glycerol drop is sitting on a deformable surface (lubricating film), wetting ridge is associated with the drop during the entire process. The entire dewetting process can be divided into three regimes, (i) early stage: hole nucleation due to the heterogeneity present on the surface, (ii) intermediate stage: hole growth and coalescence of the holes (in case of more than one hole), and (iii) late-stage: Rayleigh instability (formation of droplets) and spreading of the glycerol drop. The dewetting phenomenon (time scale and final morphology) is found very different thicker lubricating films. Upon depositing a glycerol drop on 460 nm thick PDMS film, there is an initial thinning of the lubricant film due to the Laplace pressure of the drop, which can be seen as the fluorescence intensity underneath the drop decreases. After few minutes, holes nucleate and the lubricating film is detached from the contact line of the glycerol drop. Also the lubricating film underneath the glycerol drop is disconnected from the outside lubricating film, so the residual lubricating fluid remain entrapped inside the contact line of the glycerol drop. The nucleated holes grow which finally result into multiple dewetted droplets of the diameter 5-30 $\mu$m. The diameter of the dewetted PDMS drop are very different for 460 nm and 28 nm thick lubricating films as the volumes of the lubricating fluid present underneath glycerol drop in 460 nm film is large compare to that of 28 nm film. For a lubricating film with thickness 1 $\mu$m, the initial thinning continues up to 15 minutes as seen with the decreasing fluorescence intensity. Subsequently the thinned film dewets into small PDMS droplets. In this case, the spreading of glycerol drop was much smaller during and after the dewetting of the underneath lubricating film. This is due to the fact that the wetting ridge associated with the glycerol drop is very large and prevents the glycerol drop from spreading. For 8 $\mu$m thick lubricating film, initial thinning of the underneath lubricating film continues upto about 30 minutes and the fist hole nucleates only after that. As a result, the entire dewetting takes about 50 minutes to complete. Again, due to very large wetting ridge associated with the glycerin drop, the drop does not spread much on the surface. It is interesting to note that the size of dewetted droplets for lubricant films thicker than 460 nm are very similar. This is because the final thickness of the lubricant film after initial thinning is very similar for all of them, however the the time corresponding to the lubricant thinning and the first hole nucleation is very different for all of them. Also, here is clear difference between the dewetting of the ultra-thin film to micrometer thick film. Nucleation time corresponding to first hole, increases non-linearly with film thickness. For the film having the thickness greater than few nm ($>$ 100 nm) the size of the dewetted droplets is same, but the contact area of the glycerol drop with surface is different. 

The stability of thin lubricating films of slippery surfaces can be depends on the combined effect due to the short range interaction (surface/interfacial tension) and the long range interaction (dispersion and polar) between two the phases (solid and the bounding media). In the present case, the former is defined in terms of the spreading coefficient of PDMS on glycerol; $S=\gamma_{\mathrm{DS}}-\gamma_{\mathrm{LD}}-\gamma_{\mathrm{LS}}\,=\,S_{\mathrm{LW}}\,+\,S_{\mathrm{P}}$, where, $\gamma_{ij}$, is the interfacial energy between the $i^{\mathrm{th}}$ and $j^{\mathrm{th}}$ phase, $\mathrm{D, L, S}$ represents drop, lubricating fluid, and solid phases respectively, LW represents Lifshitz-van der Waal, and P represents the polar (acid-base) interactions. For an apolar system (either bounding media is polar and the substrate is apolar or bounding media is apolar and the substrate is polar) the $S_{\mathrm{P}}=0$, hence the stability is only determined by $S_{\mathrm{LW}}$ (Lifshitz-van der Waal) interaction. Defining the stability criteria in the presence of a polar system (i.e. $S_{\mathrm{P}}\neq0$) is relatively complicated \cite{sharma1993relationship}. 

Spreading coefficient for PDMS as lubricating fluid and air as bounding medium is positive, but upon changing the bounding medium to glycerol, the spreading coefficient becomes -79.92 mN/m. A system with bounding medium as air falls into the category of the apolar system, and $S_{\mathrm{LW}}>0$ determines that a PDMS film will be stable in air. Spreading coefficient for the final system, Si/SiO$_2$/PDMS/glycerol, $S<0$ and with components $S_{\mathrm{LW}}<0$ and $S_{\mathrm{P}}<0$. Since both components of the total spreading coefficient are of the same negative sign, rupturing of lubricating PDMS films is guaranteed \cite{sharma1993equilibrium}. Latter is defined by the total free energy of interaction (per unit area) of the two bulk phases (substrate and bounding medium) separated by a PDMS lubricating film. Therefore the effective free energy is the sum of the long-range van der Waal interaction (non-retarded Lifshitz-van der Waals interaction) and relatively short-range Born interaction \cite{seemann2001dewetting}. Since, PDMS is non-polar in nature, but due to the polar nature of test liquid and substrate (bounding medium), an acid-base interaction term is added and the modified relation can be expressed as 

\begin{equation}
\label{eq:4}
\Delta{G}(\,h)=\Delta{G}_{\mathrm{LW}}(\,h)\,+\Delta{G}_{\mathrm{AB}}(\,h)\,+\frac{c}{h^8}
\end{equation}
where, $h$ is the thickness of the PDMS lubricating film and $c$ is the strength of the short-range interaction having value $1.8\times10^{-77}\,\mathrm{Jm^{6}}$ for Si wafer with the $1.7\,\mathrm{nm}$ thick silicon dioxide layer \cite{seemann2001gaining}. The first term is the contribution of non retarded Lifshitz-van der Waal for the system Si/SiO$_2$/PDMS/Glycerol and its dependence on the macroscopic parameter is 

\begin{equation}
\begin{split}
\label{eq:5}
\Delta{G}_{\mathrm{LW}}(\,h)\,=-\frac{A_{\mathrm{SiO_{2}/PDMS/glycerol\,or\, air}}}{12 \pi h^{2}}\\+\frac{A_{\mathrm{SiO_2/PDMS/glycerol\,or\, air}}-A_{\mathrm{Si/PDMS/glycerol\,or\, air}}}{12 \pi (\,{h+d_{\mathrm{SiO_2}}})\,^{2}}
\end{split}
\end{equation}
where, $A_\mathrm{SiO_2/PDMS/glycerol\,or\, air}$ and $A_\mathrm{Si/PDMS/glycerol\,or\, air}$ is the effective Hamaker constant for a three-layer system. The short-range polar/acid-base component shows the exponential dependency on film thickness \cite{sharma1993relationship},

\begin{equation}
\label{eq:7}
\Delta{G}_{\mathrm{AB}}(\,h)\,=S_{\mathrm{P}}\,\mathrm{exp}(\,\frac{{d_{\mathrm{min}}}-h}{l})\,	
\end{equation}

\begin{figure}[htbp]
	\centering
		\includegraphics[width=0.45\textwidth]{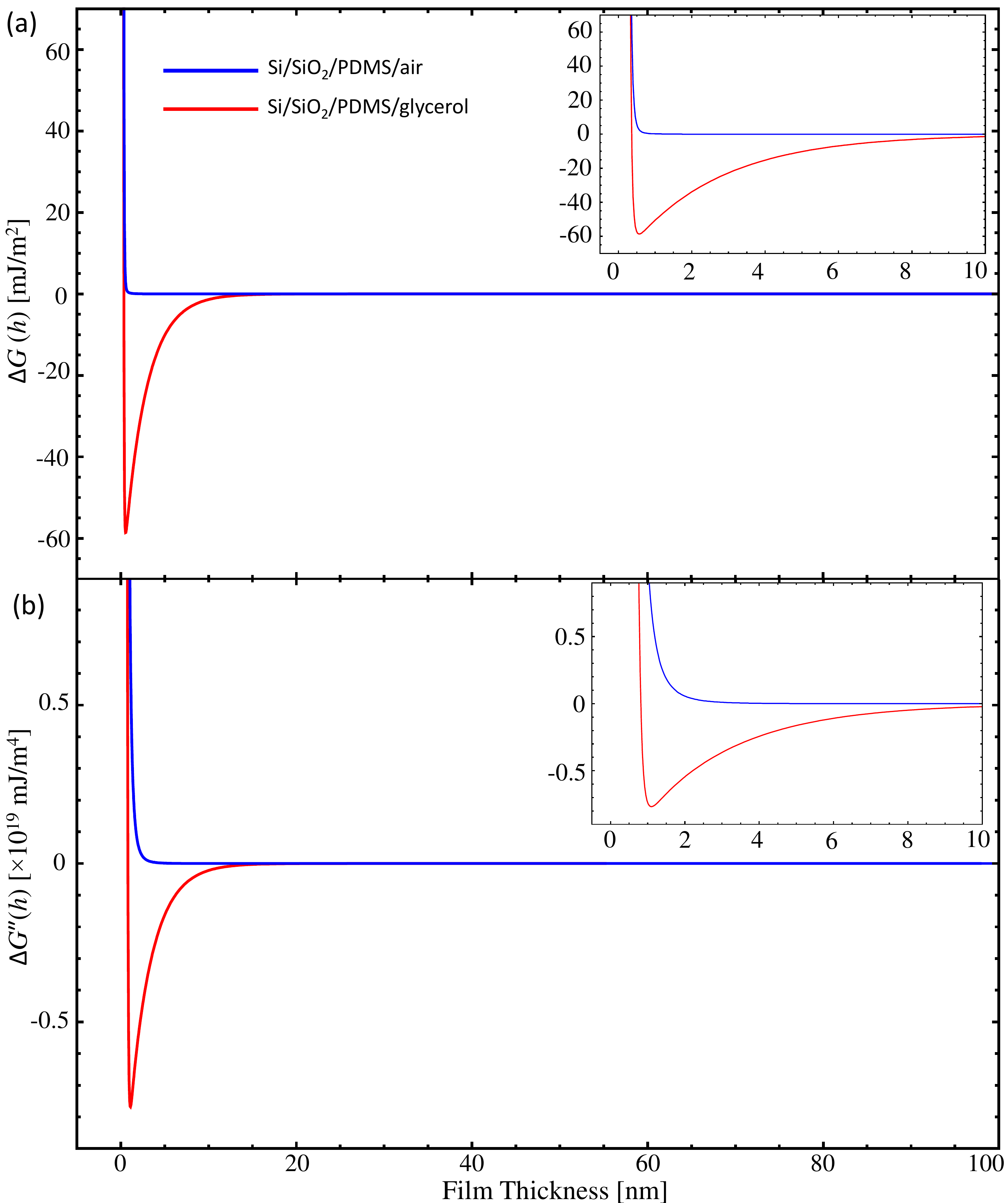}
	\caption{(a) Total excess free energy of the system Si/SiO$_2$/PDMS/Glycerol or air or water as a function of the film thickness, and (b) the second derivative of the total excess free energy of the system}
	\label{Fig:2}
\end{figure} 

The value of $d_\mathrm{min}$ for the present system as obtained from the above relation come to $0.15\,\mathrm{nm}$ whereas the correlation length for the polymer system is $l=2.5\,\mathrm{nm}$ \cite{sharma1998pattern}. Its second derivative of the total interaction energy is plotted in Fig. 2(b). Since the second derivative of the total free energy is negative with respect to all film thickness, it confirms that the underneath film is always unstable.

\begin{figure}[htbp]
	\centering
		\includegraphics[width=0.40\textwidth]{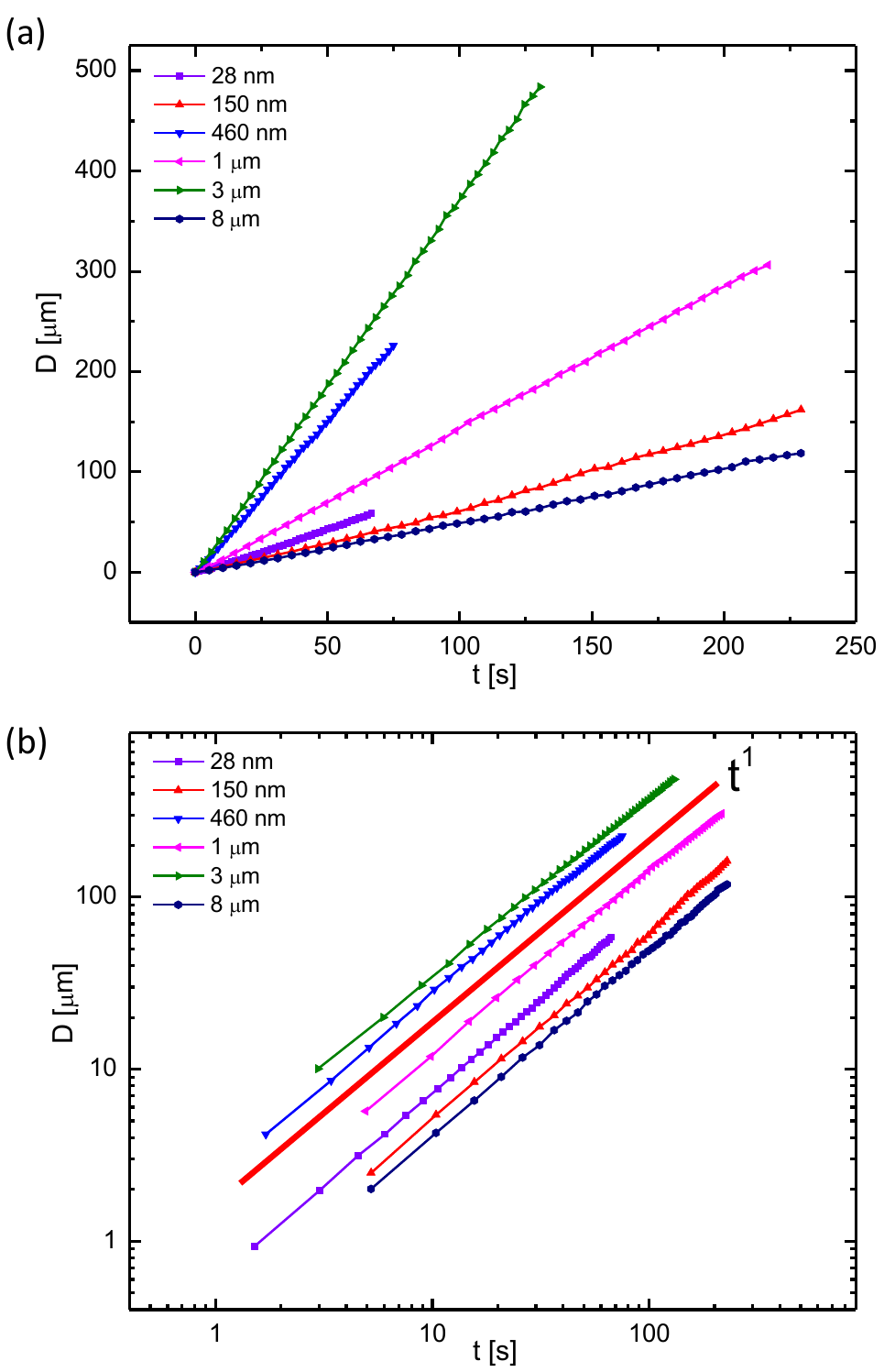}
	\caption{Dewetting distance of the PDMS film underneath the glycerol drop, (a) linear plot of dewetting distance with time, black line shows the linear fit to the data; (b) log-log plot of the D with the time and red line guides to the linear fit.}
	\label{Fig:3}
\end{figure}

The dewetting dynamics of the hole present in underneath film depend on driving force and friction forces. The driving force for the dewetting completely explains in terms of the spreading parameter. Another force that acts on this dewetting of the PDMS under glycerol is the friction force that is either due to the viscous dissipation or due to the interaction between PDMS and Si wafer (slip \cite{baumchen2009slip}). The Reynolds number $Re =\mathrm{\rho v h\,/\eta}$, where $\rho$, $v$ and $\eta$ is density of the lubricating liquid, velocity of the moving front and dynamic viscosity of the lubricating fluid is $10^{-9}$, we can neglect the inertial effects in dewetting. Only the viscous dissipation is controlling the dewetting of the film. For all thicknesses, the dewetted distance comes linearly with time [Fig. 3(a) and 3(b)]. The linear variation shows that lubricating film on silicon surface has no slip boundary condition and the breakup of the rim between two holes gives the satellite and sub-satellite drop that cannot be observed in the system that contains slip. For all the systems (no-slip or slip) the breakup of the ribbons is accompanied by the Rayleigh Plateau instability with the same wavelength of capillary waves. Dewetted distance taken up to the distance where the rim of the two holes is far away. Morphology of the final pattern is controlled by the slip present on the system and the size is depends on the volume conservation. Capillary number for the system is $Ca =\mathrm{v\eta/\gamma_{ow}}=10^{-4}$, which shows that system shows the no-slip boundary condition. In the no-slip system, the dissipation is only due to the viscous friction present in the liquid so that the dissipation is only due to the viscosity of the PDMS. For the smallest thickness of $28\,\mathrm{nm}$, droplets of the lubricating fluid were of the order of $1-2\,\mu\mathrm{m}$. While for the film thickness $>29\,\mathrm{nm}$, it is of the order of $20-30\,\mu\mathrm{m}$ due to the more viscous drag of the lubricating film when the film thickness was different. 

\begin{figure*}[htbp]
	\centering
		\includegraphics[width=1.00\textwidth]{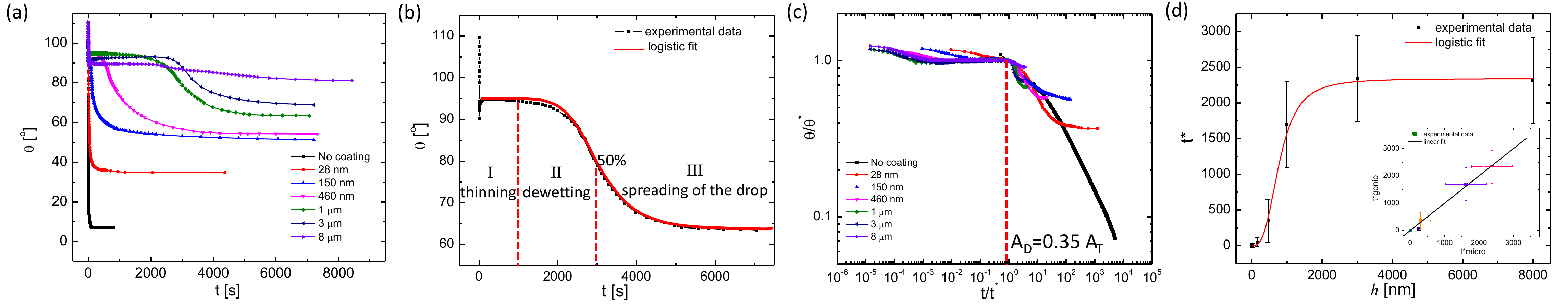}
		\caption{Effect of dewetted film on apparent contact angle; (a) characteristic of apparent contact angle with time on smooth silicon wafer for film thickness 28 nm to 8 $\mu$m; (b) scaling behavior of apparent contact angle with time for 1 $\mu$m thick film, dotted lines shows the boundary between regions; (c) log-log plot of the apparent contact angle shows the transition point at t = t$^\star$; (d) transition time with film thickness (inset shows the transition time t$^\star$ extracted from two different experiments, one top view and one side view of the glycerol drop), color data points in inset corresponds to different film thicknesses}
	\label{Fig:4}
\end{figure*}

The macroscopic behavior of the glycerol drop is affected by the intercalating film present between a drop and substrate [Fig.3(a)]. In a dry surface, the spreading coefficient of the drop is zero, that means the surface energy of the surface is large compare to the sum of the surface energy of the liquid and the interfacial tension between them. The contact angle of the drop decreases from $110^{\circ}$ to $\sim 4.2^o$ in 100 s time, due to the attractive intermolecular interaction between drop and the surface. There is sudden decrease in contact angle value, but when the same drop deposited on a PDMS coated surface, the contact angle dynamics is completely different. Due to the attractive intermolecular interaction between drop and substrate, the 28 nm thick film, dewets just after glycerol drop touches the lubricating film. Now glycerol drop is in contact with the dewetted film and high energy surface. The contact angle of the drop decrease to $42^{\circ}$ up to that time scale dewetting is already completed, after that due to the spreading of the drop on high surface energy solid surface contact angle further decreases and reaches a value $34^{\circ}$. The wetting ridge height for 28 nm thick film is in nm, so it cannot be visible in optical images and the dissipation due to the wetting ridge is quite small for 28 nm thick film. For 150 nm, thick film the apparent contact angle is $51^{\circ}$, shows the work done required to remove the lubricating film is larger compare to 28 nm film. Similar process observed for the 460 nm thick film. But going towards the micrometer thick film, due to the initial thinning then rupturing process the wetting ridge grows three times with respect to the initial wetting ridge height after 100 minutes. Due to the wetting ridge, drop cannot spread further, results the larger apparent contact angle. For 8 mm thick film, the apparent contact angle value is $\sim 81^{\circ}$. The change in the contact angle value with time follows the same scaling behavior for all thicknesses, only the limiting (equilibrium) contact angle values are different. The change in the contact angle of the glycerol drop is fitted by the relation $\theta= \frac{{A_{\mathrm{1}}}- A_{\mathrm{2}}}{1+{\frac{x}{x0}}}+A_{\mathrm{2}}$, where $A_1$ and $A_2$ are the starting point of the contact angle and limiting value of final contact angle, respectively. $X_0$ is the point where the contact angle curve change concave to the convex shape and the total change in the value of contact angle is $50 \%$ of the response. The fitting is applicable from the starting of the plateau region. We divided the contact angle behavior into three region depending upon the dewetting fraction of the film underneath the drop.
(i)	Just after the drop comes in contact with the lubricating film the drop contact angle decreases in few seconds from $110^{\circ}$ to $95^{\circ}$, this the initial settling (kinetic energy) of the drop. 
(ii)	Due to the Laplace pressure of the drop the lubricant underneath the drop squeezes lubricating fluid and the contact angle follows the linear relation with time (contact angle is independent of the film thickness underneath the drop). Laplace pressure stabilizes the film up to the value when reaches pressure difference between the drop and the film comes into an equilibrium. Since the film is continuously thinning then if there is any impurity present on the film or in substrate it feels the van der Waal interaction and creates a dry patch (hole) in the film. This nucleated hole size is in few-microns, so the contact angle of the drop slowly decreases. 
(iii)	When the dewetted distance of the film reaches to the value $35\,\%$ of the total drop diameter then the contact angle of the drop decreases exponential with time because drop is sitting on the $35\,\%$ area fraction of high energy surface (hydrophilic region). When the dewetting of the film is completed then the contact angle value decreases to the $50\,\% $ of the plateau region.
(iv) After the complete dewetting drop sits on a complete high energy surface (only small droplets of PDMS remains), so that drop must retain its equilibrium contact angle $(\sim 4^o)$. The drop is spreading on the surface, results further decrease in contact angle. Due to the presence of wetting ridge reduces the possibility to attain the equilibrium angle (due to the viscous drag). 

The scaled value of the contact angle [Fig. 4(c)] shows the transition in contact angle value from linear to exponential decay at time when the dewetted film area is the $35\,\%$ of the total area $(A_D=0.35\,A_T)$. This shows that contact angle value starts changing only when sufficient amount of film underneath the drop dewets. The apparent contact angle value is increases with the film thickness, clearly reflects the effect of viscous dissipation during spreading of the drop. The $t^\star$ value changes exponentially with the thickness initially up to $2\,\mu\mathrm{m}$ film then it saturates [Fig. 4(d)]. The $2\,\mu\mathrm{m}$ film thickness value is the transition thickness after that the transition time $t^\star$ is independent of the film thickness that indicates that the initial thinning process is due to Laplace pressure of the drop and later time it depend on the long-range interaction.  Below this thickness, it increases exponentially gives the idea that both Laplace pressure and Dissjoining pressure acts across the interfaces at the same time, which results the dewetting.   
The characteristic time measured from two independent experiments (side view optical images or from top view microscopic images) follows the linear relation [Fig. 4(d) inset]. 

In summary, we have investigated the dewetting of lubricating films underneath aqueous drops on lubricated slippery surfaces for different thicknesses of lubricating films. On slippery surfaces with high surface energy solids, lubricating films are not stable due to the long-range van der Waals interaction. As aqueous drops are deposited on lubricant coated surfaces, the underneath lubricant films are initially thinned due to the Laplace pressure of the drop, which subsequently undergo dewetting. The characteristic dewetting data shows scaling for different thicknesses with critical contact angle and time corresponding to the $35\,\%$ dewetting of lubricant films. Hole growth during dewetting (measured as dewetting distance) shows linear behavior with time indicating the no-slip boundary condition for the system. 

\nocite{*}

\bibliography{main}

\end{document}